\documentclass[useAMS,usenatbib,usegraphicx]{mn2e}
\usepackage{apjfonts}
\usepackage{journals}

\usepackage{amssymb}  

\newcommand\heii{\hbox{He\,{\sc ii}}}
\newcommand\hei{\hbox{He\,{\sc i}}}
\newcommand\hi{\hbox{H\,{\sc i}}}
\newcommand\teff{\mbox{$T_{\rm eff}$}}
\newcommand{\bvfreq}{Brunt-V\"ais\"al\"a frequency}

\title{White dwarf envelopes: further results of a non-local model of
  convection}

\author[Montgomery \& Kupka]
       {M. H. Montgomery$^1$ \& F. Kupka$^{2,3}$ \\
       $^1$Institute of Astronomy, University of Cambridge, Madingley Road,
       Cambridge CB3 0HA, United Kingdom \\
       $^{2}$Astronomy Unit, School of Mathematical Sciences, Queen Mary, 
       University of London, Mile End Road, London E1 4NS \\
       $^{3}$Max-Planck-Institute for Astrophysics, Karl-Schwarzschild-Str.~1,
       85741 Garching, Germany}

\date{
  Accepted 2004 January 22}

\pagerange{\pageref{firstpage}--\pageref{lastpage}}
\pubyear{2004}

\begin{document}

\maketitle

\label{firstpage}

\begin{abstract}

We present results of a fully non-local model of convection for white
dwarf envelopes. We show that this model is able to reproduce the
results of numerical simulations for convective efficiencies ranging
from very inefficient to moderately efficient; this agreement is made
more impressive given that \emph{no closure parameters have been
adjusted} in going from the previously reported case of A-stars to the
present case of white dwarfs; for comparison, in order to match the
peak convective flux found in numerical simulations for both the white
dwarf envelopes discussed in this paper and the A-star envelopes
discussed in our previous work requires changing the mixing length
parameter of commonly used local models by a factor of 4. We also
examine in detail the overshooting at the base of the convection zone,
both in terms of the convective flux and in terms of the velocity
field: we find that the flux overshoots by $\sim 1.25\, H_P$ and the
velocity by $\sim 2.5\, H_P$.  Due to the large amount of overshooting
found at the base of the convection zone the new model predicts the
mixed region of white dwarf envelopes to contain at least 10 times more
mass than local mixing length theory (MLT) models having similar
photospheric temperature structures.  This result is consistent with
the upper limit given by numerical simulations which predict an even
larger amount of mass to be mixed by convective overshooting.  Finally,
we attempt to parametrise some of our results in terms of local
MLT-based models, insofar as is possible given the limitations of MLT.

\end{abstract}


\section{Introduction}

Understanding white dwarf stars is crucial for many areas of
astrophysics. First, their masses can be used to place constraints on
mass loss in the post-Main Sequence phase and hence on the
initial-final mass relation \citep{Weidemann00}. Second, their
temperatures can be used to derive ages, either individually, for
clusters, or for the local Galactic disk
\citep[e.g.][]{Winget87,Wood92}. Complementary to this, white dwarfs
are observed to pulsate in specific temperature ranges, and these
pulsations allow us to probe and constrain the interior structure of
these stars \citep{Winget98}. Through their pulsations, we can use
asteroseismology to examine various physical processes such as nuclear
reaction rates \citep{Metcalfe03}, chemical diffusion
\citep{Montgomery01}, crystallisation \citep{Winget97,Montgomery99a},
and neutrino emission \citep{Obrien98}.

In addition, these pulsations are most likely driven through their
interaction with the surface convection zone in these stars
\citep{Brickhill91a,Wu97,Goldreich99a}, meaning that the onset of
pulsations (in terms of \teff) is linked with the convection zone
reaching a certain depth. In many cases the observed amplitudes imply
that the depth and structure of these convection zones should vary
appreciably (by a factor of several in mass) during a pulsational
cycle, and this is thought to be the origin of the dominant
nonlinearities in many of the observed lightcurves
\citep{Brickhill92b,Wu01}. Given the simplification afforded by the
separation of the convective turnover timescale ($\sim$~1~s) and the
pulsation timescale ($\sim$~100~s), it is possible to use the
pulsations themselves to sample and constrain the convection zones of
these stars \citep[ Montgomery, in preparation]{Ising01}.
Consequently, other than the well-studied convection zone of the Sun,
white dwarfs may offer the best chance for testing theories of stellar
convection.

In this paper, we apply the Reynolds stress model formalism for convection
\citep[see][]{Kupka02} to envelope models of both DA (hydrogen) and DB
(helium) white dwarfs. As was the case for convection in A-stars, this
problem is made easier numerically by the fact that the white dwarf
models we consider have relatively thin surface convection zones, and
hence shorter thermal relaxation timescales. As a consequence, we have
treated models in which convection is not the dominant form of energy
transport, i.e., $F_C/F_T\la 0.5$; in this regime, convection is much
more sensitive to details in the modelling.

\begin{figure*}
\includegraphics[height={1.00\columnwidth},angle=-90]{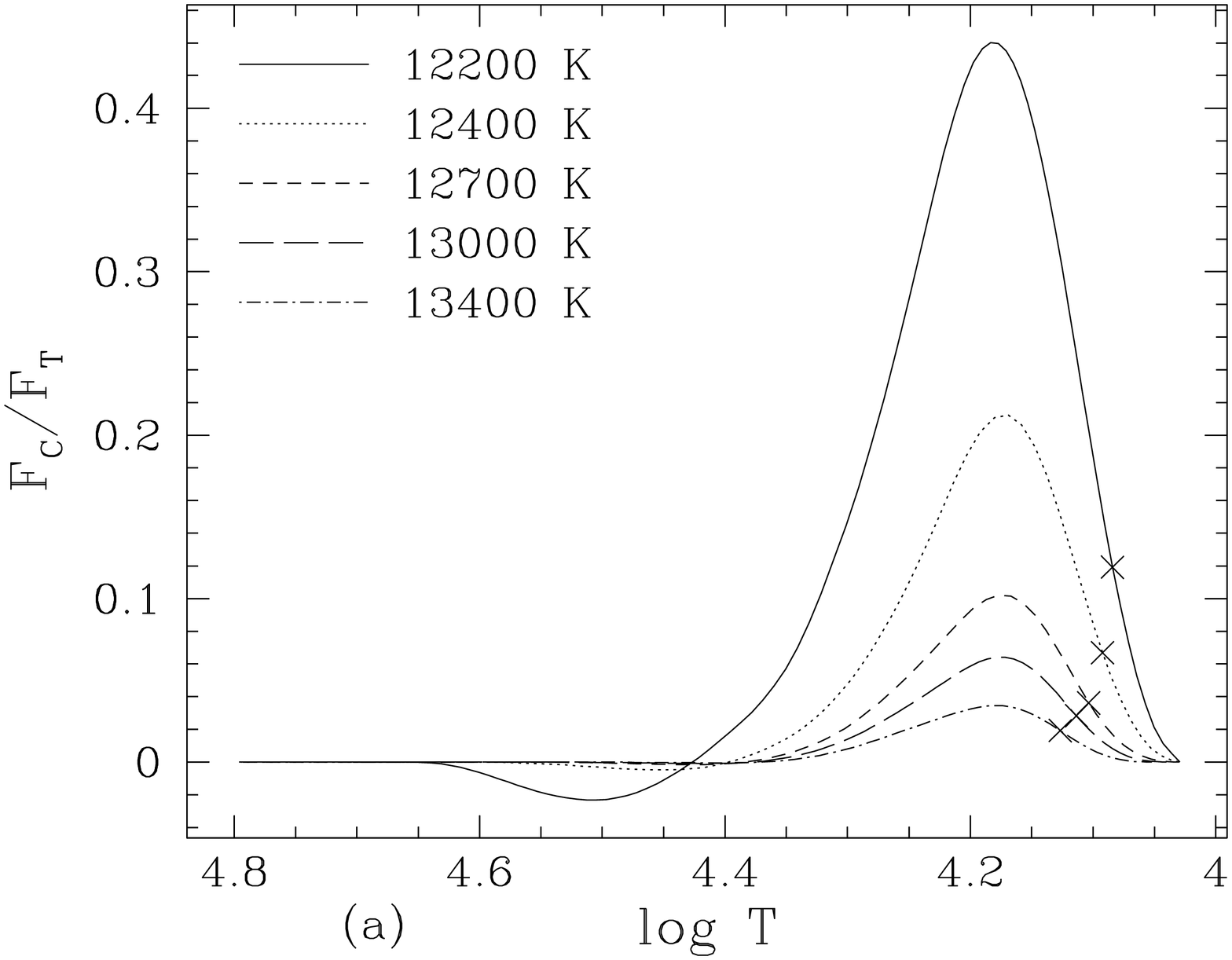}
\hfil
\hspace{2.5em}
\includegraphics[height={1.00\columnwidth},angle=-90]{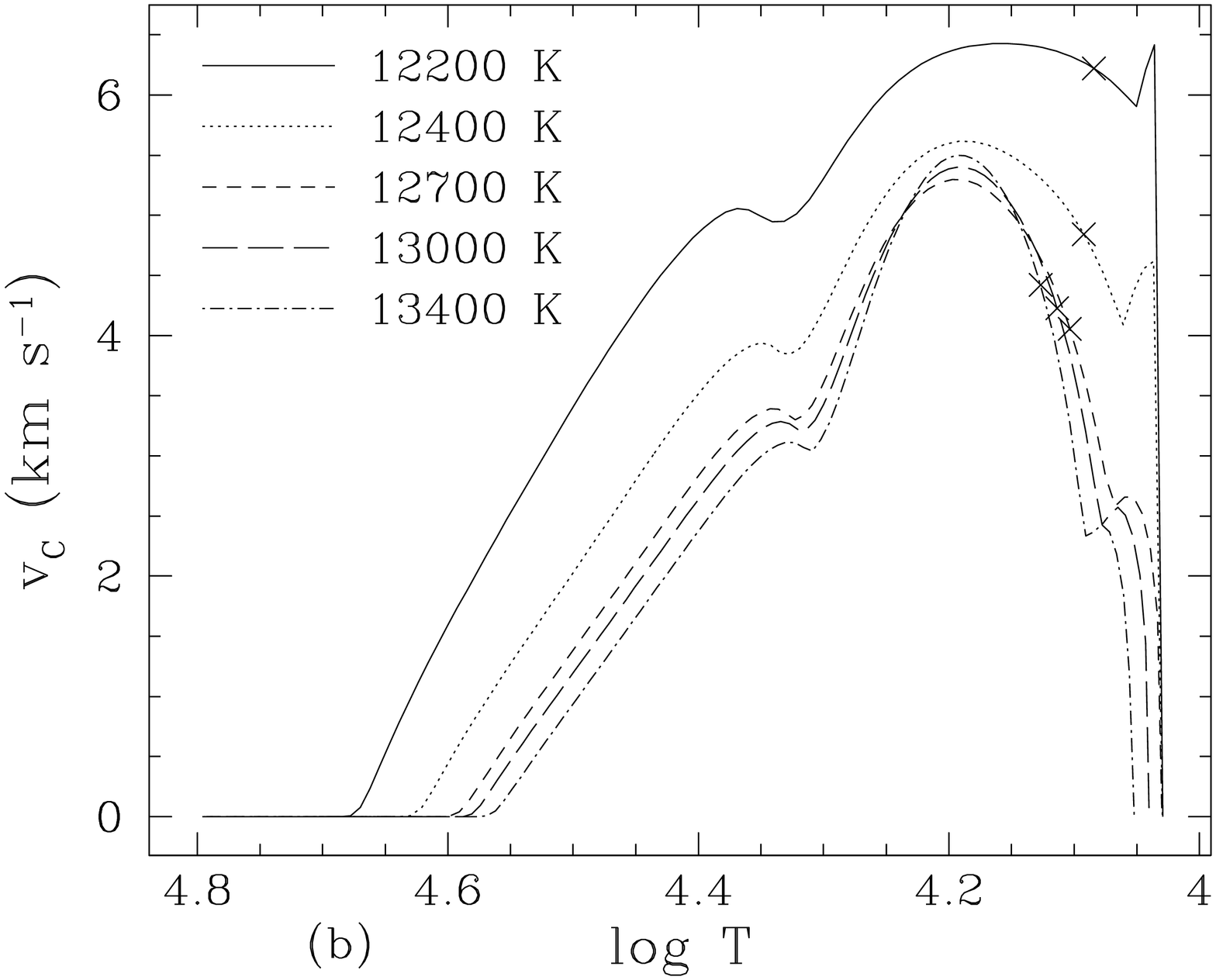}
\caption{(a) The fraction of the flux carried by convection for five 
  DA white dwarf models with the indicated effective temperatures,
  where we have taken $\log T$ as our radial variable; $\log g = 8.0$
  and $X = 1.00$ ($Z = 0.00$) for all models. The cross on each curve
  near $\log T \sim 4.1$ shows the location where $\tau=2/3$ for each
  model.  As expected, convection becomes more dominant with
  decreasing \teff. (b) The same as (a) but for the rms of the
  vertical component of the convective velocity.
\label{models}
}
\end{figure*}

In the sections which follow, we give a brief outline of the physics
and the numerical procedure used to compute our envelope models.
Results are then presented for models of DA and DB white dwarfs, as a
function of \teff. Finally, where possible we compare our results to
those of numerical simulations and to models used for fitting white
dwarf spectra. We show that our calculations qualitatively follow the
simulations, especially in terms of the convective flux, and that the
photospheric temperature structure of our models is similar to that of
previous investigations.

\section{Description of Model}   \label{Sect2}

\begin{table*}
\centering
\begin{minipage}{1.00\textwidth}
  \caption{Convection zone parameters of DA white dwarf models obtained 
    with the non-local model. The overshooting (OV) is defined as the
    distance in pressure scale heights from the minimum of $F_{\rm
      C}/F_{\rm T}$ to the point where $|F_{\rm C}/F_{\rm T}|\sim
    10^{-6}$ (similar to $L_{\rm t}$ in \citealt{Zahn91}). OV[mix] is
    the velocity overshooting; it is defined as the distance from the
    base of the formally unstable convective region ($\nabla \sim
    \nabla_{\rm ad}$) to the point where a linear extrapolation of the
    velocity vanishes (see Fig.~\protect\ref{mltcomp}). Finally,
    $\alpha_{\rm eff}$ is the value of $\alpha$ for the given mixing
    length theory (either ML1 or ML2) required to reproduce the peak
    flux in the convection zone.}
\setlength{\tabcolsep}{5.62pt}
  \begin{tabular}{cccccccccccc}
    \hline
    \teff
 &  $\log g$ 
 & $\log(M_{\rm CZ}/M_\star)$
 & $(F_{\rm C}/F_{\rm T})_{\rm max}$ 
 & OV 
 & $(v_{\rm C})_{\rm max}$ 
 & $(v_{\rm C})_{\tau=2/3}$ 
 & OV[mix] 
 & $(p_{\rm turb}/p_{\rm tot})_{\rm max}$ 
 & $(p_{\rm turb}/p_{\rm tot})_{\tau=2/3}$ 
 & \multicolumn{2}{c}{$\alpha_{\rm eff}$} \\
 (K)  & & & & (in $H_p$) & (km~s$^{-1}$) & (km~s$^{-1}$) & (in $H_p$)  & 
 & & (ML1) & (ML2) \\
    \hline  \\[-0.9em]
12200 & 8.00 &  -14.72 & 0.4402 & 1.41 & 6.43 & 6.22 & 2.75 & 0.290 & 0.242 & 1.66 & 0.65 \\ 
12400 & 8.00 &  -14.93 & 0.2124 & 1.37 & 5.61 & 4.84 & 2.45 & 0.172 & 0.155 & 1.57 & 0.64 \\ 
12700 & 8.00 &  -15.07 & 0.1020 & 1.24 & 5.30 & 4.05 & 2.22 & 0.125 & 0.106 & 1.73 & 0.72 \\ 
13000 & 8.00 &  -15.14 & 0.0642 & 1.21 & 5.40 & 4.22 & 2.19 & 0.125 & 0.107 & 2.04 & 0.85 \\ 
13400 & 8.00 &  -15.22 & 0.0346 & 1.14 & 5.50 & 4.41 & 2.15 & 0.125 & 0.106 & 2.53 & 1.06 \\[0.5em] 
12700 & 8.30 &  -15.28 & 0.3414 & 1.29 & 5.58 & 4.98 & 2.50 & 0.184 & 0.161 & 1.54 & 0.62 \\ 
12700 & 7.70 &  -14.70 & 0.0473 & 1.23 & 5.88 & 4.91 & 2.29 & 0.152 & 0.141 & 2.30 & 0.96 \\ 
    \hline
\label{params}
\end{tabular}
\end{minipage}
\end{table*}

The convection model used here is an extension of the \citet[
hereafter CD98]{Canuto98} model which requires the solution of five
differential equations of first order in time and second order in
space for the hydrodynamic moments $K$, $\overline{\theta^2}$,
$J=\overline{w\theta}$~$=F_C/(\rho c_p)$, $\overline{w^2}$, and
$\epsilon$, and of an additional equation for the time evolution of
$T$ \citep[cf.\ equations (1)--(5) and (8) in ][]{Kupka99}; here and
throughout this paper, $K$ is the turbulent kinetic energy, $\theta$
and $w$ are the temperature and velocity fluctuations, respectively,
and $\epsilon$ is the turbulent kinetic energy dissipation rate. This
system is completed by an equation for the total pressure
(`hydrostatic equilibrium' including turbulent pressure, equation (7)
in Kupka 1999b) and for the mass (`conservation of mass'). We solve
this set of differential equations on an unequally spaced mass grid,
with the zoning chosen so as to resolve the gradients in the various
quantities. The model equations were derived in a series of papers by
\citet{Canuto92,Canuto93}, CD98, and \citet{Canuto01}.  Compressibility
effects are taken into account following \citet{Canuto93}. The
adoption and solution of these equations for stellar envelopes 
is described in \citet{Kupka02}, and it is this model which we apply
here as well.  The closure parameters for correlations such as
$\overline{\theta \, \partial p'/ \partial x_j}$, i.e. between
fluctuations of the temperature and the pressure gradient, which
appear in the non-local model, have been taken over from the previous
paper \citep[see][]{Kupka02}. We note here that no mixing length is
used in this model due to a separate evolution equation for the
turbulent kinetic energy dissipation rate $\epsilon$.

For our calculations, we have used a Prandtl number of $10^{-6}$ as a
typical value for the outer part of white dwarf envelopes; values 2
orders of magnitude smaller than this produce essentially identical
results.  We note that values 2 orders of magnitude {\em larger}
(i.e., $10^{-4}$) can alter our results for the hottest models at up
to the $\sim\,$10\% level. For the constitutive physics, we use the
equation of state and opacity data from the OPAL project
\citep{Rogers96,Iglesias96}.  Since we are treating white dwarfs with
pure or nearly pure surface layers, we have taken the metallicity to
be zero ($Z=0.0$), with $X=1.0$ for the DA (hydrogen spectrum) white
dwarfs and $Y=1.0$ for the DB (helium spectrum) white dwarfs.

While the convection zones we have treated here and in \cite{Kupka02}
are relatively thin surface convection zones, we have solved the full
equations for spherical geometry. For upper (outer) boundary
conditions, we fix the temperature, gravity, and stellar radius to be
equal to those obtained from an envelope model assuming local (MLT)
convection, which itself has a given \teff, $\log g$, and $R_{\star}$;
thus, both the location of the model in the H-R diagram and its mass
are specified (these may be taken either from a self-consistent
stellar model or freely specified). We mention that the outer
photospheres of these envelope models are completely radiative, so the
use of MLT in these models leaves no direct imprint on our subsequent
solutions, with the exception of the value derived for the stellar
radius, which is slightly affected. At the lower boundary, we assume a
constant input luminosity $L_{\star}$ equal to the luminosity at the
stellar surface.  The complete system is then integrated in time
(currently by a semi-implicit method) until a stationary, thermally
relaxed state is found. The mass shells can be re-zoned to a different
relative size to resolve, e.g. steep temperature gradients that may
appear and/or disappear during convergence. The radiative envelope
below the convection zone may then be obtained from a simple downward
integration.

\section{Results}

\label{results}

In the following sections, we report results for DA (hydrogen) white
dwarfs and DB (helium) white dwarfs.  For most of our calculations we
have used the canonical value of the white dwarf surface gravity of
$\log g = 8.0$, which corresponds to a stellar mass of $\sim 0.6
M_{\odot}$, although we also present selected results for $\log g =$~7.7
and 8.3.

The temperatures we have explored in these models range from high
temperatures in which convection is very inefficient to cooler
temperatures for which the convection becomes deeper although not yet
adiabatic. Since convection in the DA's is driven by \hi\ partial
ionisation and in the DB's mainly by \heii\ partial ionisation, the
temperatures of the DB models are much higher; for our DA models, we
have chosen temperatures in the range 12,200--13,400~K, and for the DB
models 28,000--35,000~K.  Observationally, these temperatures
approximately correspond to the onset of pulsations in these stars,
the so-called `blue edge' of the pulsational instability strip, which
is $\sim\,$12,500~K for the DA's \citep{Bergeron95} and
$\sim\,$28,000~K for the DB's \citep{Beauchamp99}. These temperatures
can be explained either in terms of linear instability due to
convective driving \citep{Brickhill91a,Goldreich99a} or the
traditional $\kappa$-$\gamma$ driving mechanism
\citep{Winget82,Winget82a}. Thus, the {\em nature} of convection in
these stars has a large impact on the properties of the pulsations.

\subsection{DA models}

For the DA's, we have considered temperatures between 12,200~K and
13,400~K; Fig.~\ref{models} shows our central results.  First, the
models are all strongly convective in the photosphere, with the
convective flux being a substantial fraction of the maximum value in
the \hi\ convection zone.  Second, we see that the photospheric
velocities are even larger, attaining values at least as large as 75\%
of their maximum values within the convection zone.  Far out in the
photospheres of these models (the crosses indicate the point at which
$\tau=2/3$), we see that the models do become radiative, justifying
our use of fully radiative outer boundary conditions.  We note that
the wiggles near $\log T \sim 4.35$ and $4.07$ in the velocity field
are due to terms in the equations for third order moments, which
represent non-local transport, and whose functional form depends on the
sign of $N^2$ (the square of the \bvfreq); a smoother model for these
terms would remove this feature from the velocities.\footnote{For
  instance, the non-Gaussian closures recently suggested by
  \citet{Gryanik02} should not produce such a feature. Their new model
  promises a more realistic approach to third and fourth order moments
  in Reynolds stress models, but requires the solution of additional
  differential equations.}  The actual velocity distribution in
transition regions is expected to be smooth, and is found to be so in
numerical simulations. We also have to emphasise that the $\log T$
scale does not properly resolve the upper and mid photosphere, hence
the apparently very steep drop of velocity at the surface.
Comparisons with the convective flux displayed in Fig.~\ref{models}a
and the kinetic energy flux shown in Fig.~\ref{fkin} below indicate
what a plot as a function of logarithmic optical depth ($\log
\tau_{\rm ross}$) reveals more clearly: a steady and by no means
abrupt decay towards zero. Moving the upper boundary further outwards
by two orders of magnitude on the optical depth scale would not be
noticed in Fig.~\ref{models}b, as the photospheric temperature has
already reached a constant value there due to the underlying
approximation of grey radiative transfer.

In Table~\ref{params}, we give a summary of our results for these DA
models. For each model, we give the maximum value of the convective
flux, $(F_C/F_T)_{\rm max}$, the overshooting in pressure scale
heights of the flux, OV, and the velocity, OV[mix], and the maximum
and photospheric values of the vertical component of the convective
velocity and the turbulent pressure, $v_{\rm C}$ and $p_{\rm
  turb}/p_{\rm tot}$, respectively. We also give $\log (M_{\rm
  CZ}/M_\star)$, where $M_{\rm CZ}$ is the total mass of the
convection zone, defined as the region which is mixed (including
velocity overshooting). Finally, we list values of $\alpha_{\rm eff}$,
which is the value of $\alpha$ which a given MLT model requires in
order to reproduce the maximum convective flux.  The columns labelled
`ML1' and `ML2' denote the values obtained using two different
versions of MLT, by \citet{Bohm-Vitense58} and \citet{Bohm71},
respectively. These versions of MLT will be discussed further
in Section~\ref{mlt}.

Over the temperature range of the models, we see that for ML1
convection that $\alpha_{\rm eff}$ lies roughly in the range 1.5--2.5;
clearly, a local MLT using a single value of $\alpha$ would be unable
to reproduce these results.  At any rate, this range of values is
consistent with those found by \cite{Ludwig94} in comparison with
numerical simulations, and is also consistent with the model
atmosphere fits of \citet{Koester94}. In Section~\ref{mlt}, we make a
more systematic comparison with MLT models (in terms of both ML1 and
ML2), taking into account the photospheric temperature structure of
the models.

\begin{figure}
\includegraphics[width=\columnwidth]{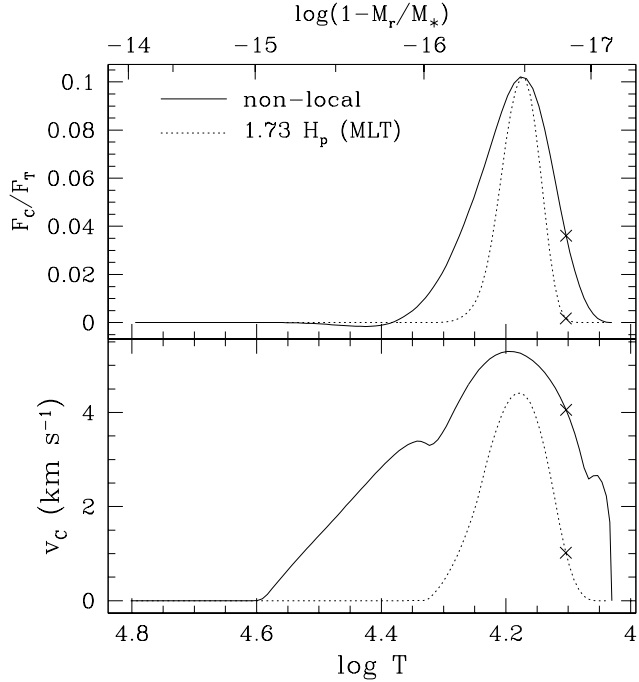}
\caption{A comparison of the convective fluxes (upper panel) and 
  velocities (lower panel) of the non-local model and of an MLT model
  ($l = 1.73 H_p$), for $\teff=12,700$~K. We see that the MLT model
  cannot simultaneously match the height and width of the convective
  flux.  In addition, the velocity field of the MLT model does not
  extend as far out into either the photosphere or downward into the
  envelope as that of the non-local model. Along the top axis we have
  indicated the depth in terms of the envelope mass $\log
  (1-M_r/M_\star)$, so that we can see the relative amounts of mass in
  the formally unstable and overshooting regions.
\label{mltcomp}
}
\end{figure}

In Fig.~\ref{mltcomp}, we make a detailed comparison between the flux
and velocity structure of our $\teff =$~12,700~K envelope model with
that obtained using the standard MLT prescription for convection.
There are at least four aspects of our convection zone solutions which
we might wish to compare to MLT: the peak flux and its overshooting
depth (i.e.\ the convectively mixed region), and the peak velocity and
\emph{its} overshooting depth. Given
the inadequacies of a local MLT, however, it is only possible to fit
one of these quantities at a time. Thus, we have adjusted the
mixing-length parameter $\alpha$ so that the two models have the same
maximum flux; this is achieved for a value of $\alpha = 1.73 H_{P}$.
In terms of the convective flux, we see that the non-local model
predicts a wider convective region with overshooting extending down to
$\log T \sim 4.53$, as compared to the local MLT model, which has its
base at $\log T \sim 4.3$.  Significantly, the non-local model also
predicts a much larger value for the photospheric flux. In terms of
the convective velocities, the differences are even larger. The
non-local model predicts velocities 50\% larger than the local model,
and due to overshooting, these velocities also extend much deeper,
down to $\log T\sim 4.6$, as compared to $\log T \sim 4.33$ for the
local model. Thus, we find that the mass of material `stirred' by the
convection zone is an order of magnitude larger than that which is
formally unstable according to a local stability criterion. Among
other things, this can have consequences for the diffusion of elements
in white dwarf envelopes, as was pointed out by \citet{Freytag95} and
\citet{Freytag96}, who, in addition, found an even larger amount of
material stirred below the convection zone in their numerical
simulations of DA envelopes (see also Section~\ref{Sec_numsim}).

\begin{figure}
\includegraphics[height=\columnwidth,angle=-90]{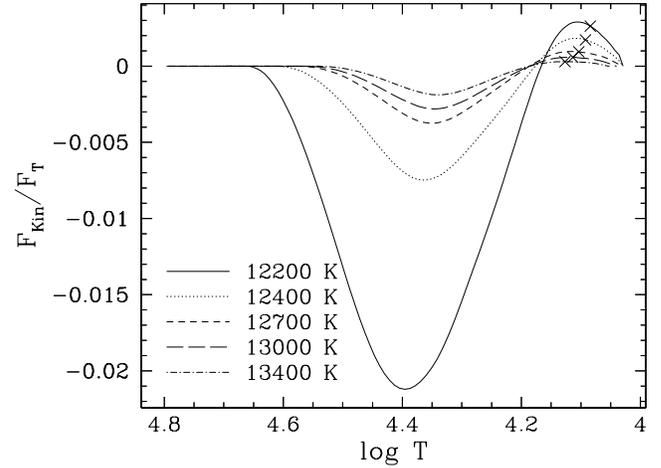}
\caption{The kinetic energy flux as a function of $\log T$, for the five
  different models shown in Fig.~\ref{models}. As expected, the cooler
  models have larger fluxes. Most significantly, however, these
  numbers show that $|F_{\rm kin}|$ is essentially negligible for the
  models we have examined.
\label{fkin}
}
\end{figure}

\subsection{Kinetic energy flux and non-locality}

In Fig.~\ref{fkin} we plot the kinetic energy flux for the five
different models shown in Fig.~\ref{models}. Besides the fact that the
cooler models have larger fluxes, which is to be expected, we see from
the magnitudes of these fluxes that $F_{\rm kin}$ is essentially
negligible for the models we have examined. Taking this result at face
value, we are thus in a different regime from that of the Sun, in
which $|F_{\rm kin}/F_{\rm T}|$ may be as large as 20 per cent
\citep[cf.\ ][]{Stein98,Kim98}.  Although a positive $F_{\rm kin}$ in
the photosphere indicates that the skewness of spectral lines (and
their NLTE cores) produced in this region should also be positive
(CD98, cf.\ the discussion in \citealt{Kupka02}), the small magnitude
of $F_{\rm kin}$ means that this effect will be small and probably
difficult to measure. In the future, it will hopefully be possible to
make observational tests of these predictions.

Although the kinetic energy flux in these models appears to be small,
this transport of kinetic energy may lead to an equilibrium state
having large velocity fields outside of the formally convective
regions, provided that the local dissipation rate of kinetic energy is
small enough. For instance, in Fig.~\ref{mltcomp} we see that both
below ($\log T\ga 4.4$) and above ($\log T\la 4.05$) the formally
convective region there is significant overshooting of the velocities.
This can be understood from Fig.~\ref{fkin} since $F_{\rm kin}$ is
positive at the top of the formally convective region ($\log T\sim
4.1$), indicating that kinetic energy is being transported outward,
while $F_{\rm kin}$ is negative at the bottom of the formally
convective region ($\log T\sim 4.4$), indicating that kinetic energy
is being transported inward.  This velocity overshooting is an
essential feature of these models, and is an effect which cannot be
captured using local MLT-type models (see Fig.~\ref{mltcomp} above).

\begin{figure*}
\includegraphics[height={\columnwidth},angle=-90]{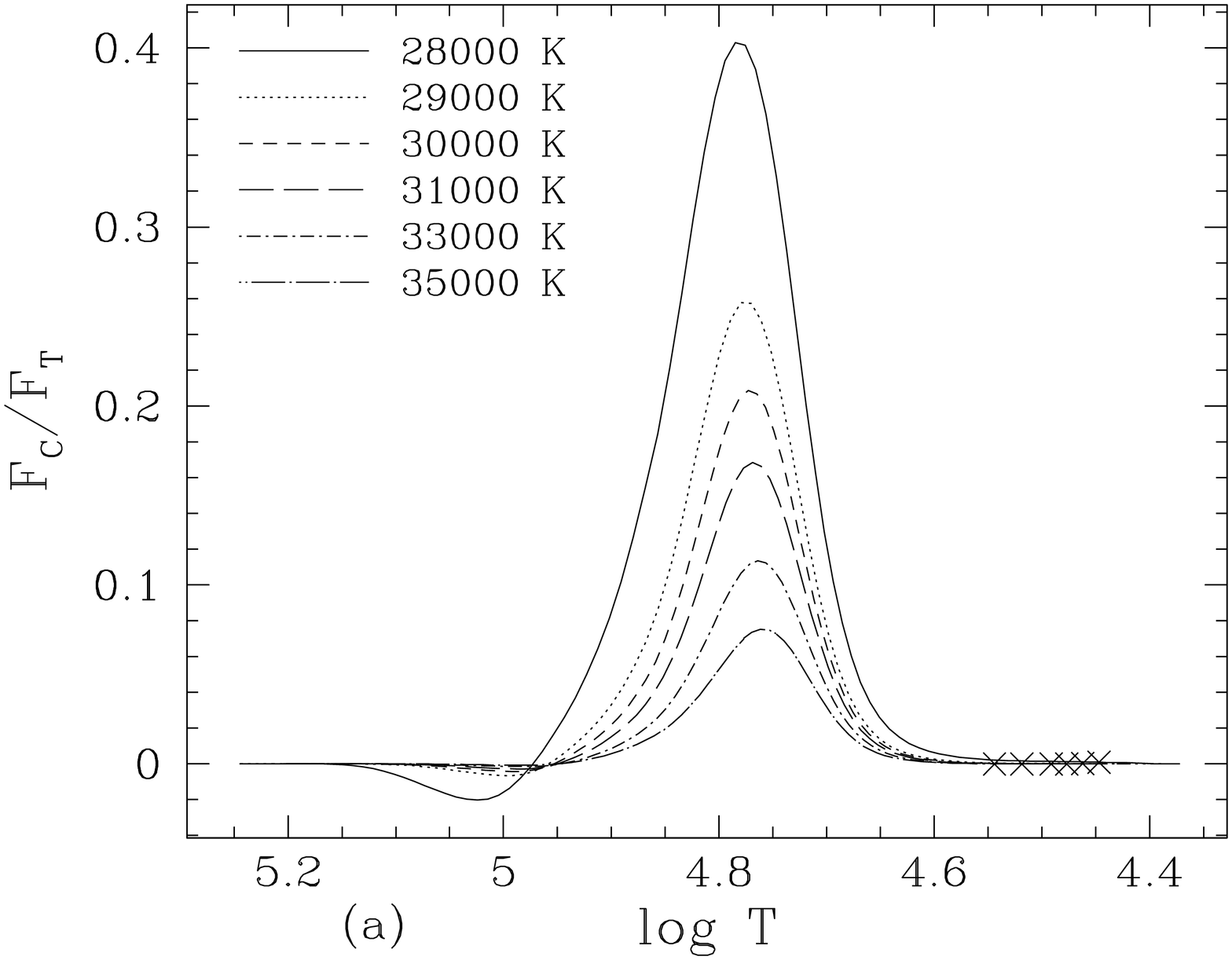}
\hfil
\hspace{2.5em}
\includegraphics[height={\columnwidth},angle=-90]{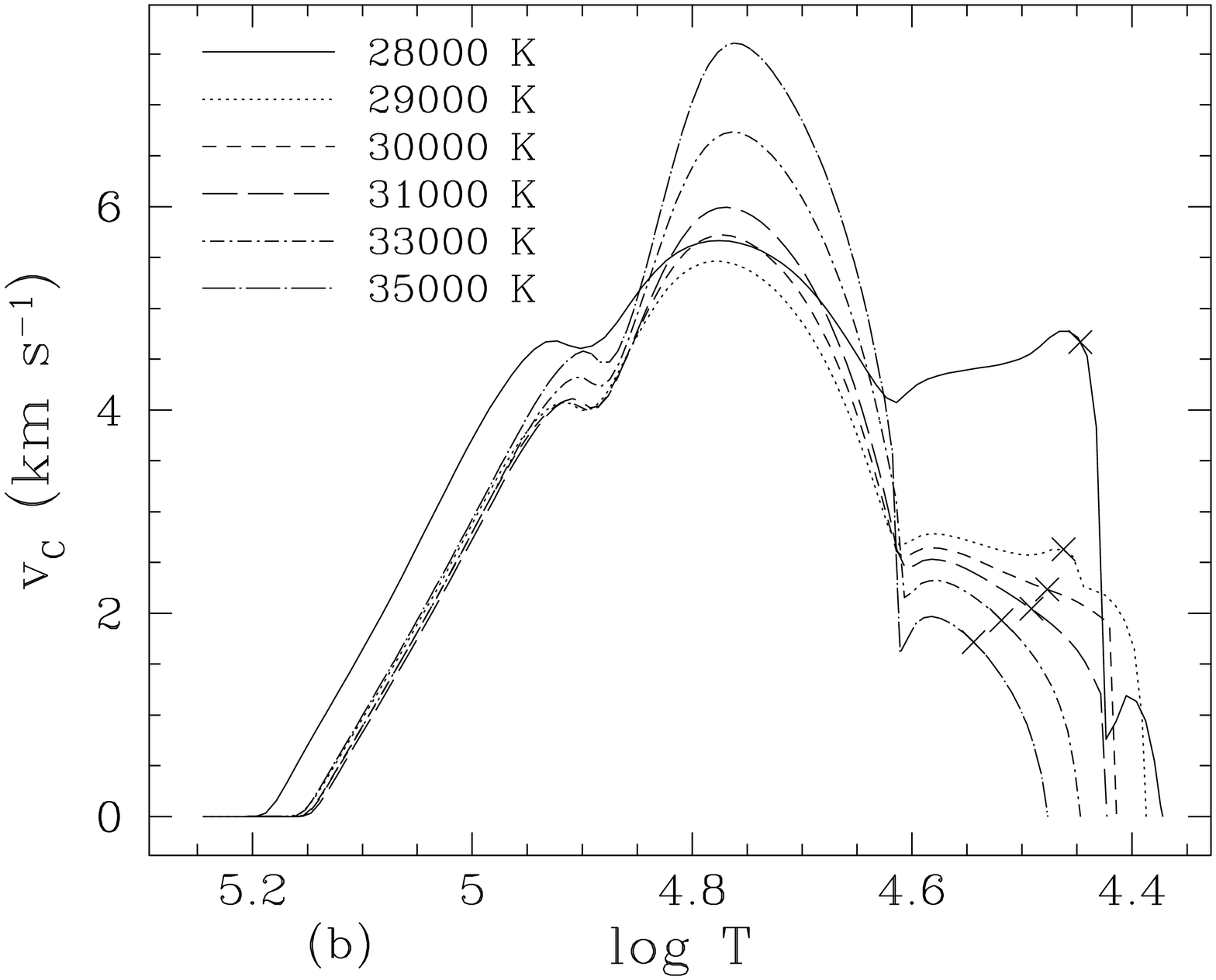}
\caption{(a) The fraction of the flux carried by convection for six 
  DB white dwarf models with the indicated effective temperatures, as
  a function of $\log T$; $\log g = 8.0$ and $Y = 1.00$ ($Z = 0.00$)
  for all models. The crosses in the range of $\log T \sim$ 4.45--4.55
  show the respective locations of the photospheres ($\tau=2/3$) of
  the different models. As expected, convection becomes more efficient
  with decreasing \teff. (b) The same as (a) but for the rms of the
  vertical component of the convective velocities.
\label{models_DB}
}
\end{figure*}

\begin{table*}
\centering
\begin{minipage}{1.00\textwidth}
  \caption{The same as Table~\protect\ref{params}, but for
    DB white dwarf models. }
\setlength{\tabcolsep}{5.62pt}
  \begin{tabular}{cccccccccccc}
    \hline
    \teff
 & $\log g$
 & $\log(M_{\rm CZ}/M_{\star})$ 
 & $(F_{\rm C}/F_{\rm T})_{\rm max}$
 & OV 
 & $(v_{\rm C})_{\rm max}$ 
 & $(v_{\rm C})_{\tau=2/3}$ 
 & OV[mix] 
 & $(p_{\rm turb}/p_{\rm tot})_{\rm max}$ 
 & $(p_{\rm turb}/p_{\rm tot})_{\tau=2/3}$ 
 & \multicolumn{2}{c}{$\alpha_{\rm eff}$} \\
 (K)  & & & & (in $H_p$) & (km~s$^{-1}$) & (km~s$^{-1}$) & (in $H_p$)  & 
 & & (ML1) & (ML2) \\
    \hline  \\[-0.9em]
28000 & 8.00 & -13.28 & 0.4030 & 1.33 & 5.67 & 4.67 & 2.53 & 0.168 & 0.166 & 1.14 & 0.45 \\ 
29000 & 8.00 & -13.42 & 0.2577 & 1.32 & 5.46 & 2.62 & 2.36 & 0.102 & 0.057 & 1.07 & 0.43 \\ 
30000 & 8.00 & -13.48 & 0.2086 & 1.28 & 5.72 & 2.24 & 2.36 & 0.111 & 0.040 & 1.12 & 0.45 \\ 
31000 & 8.00 & -13.51 & 0.1683 & 1.29 & 5.99 & 2.04 & 2.36 & 0.120 & 0.032 & 1.16 & 0.48 \\ 
33000 & 8.00 & -13.55 & 0.1136 & 1.31 & 6.73 & 1.93 & 2.47 & 0.147 & 0.027 & 1.28 & 0.53 \\ 
35000 & 8.00 & -13.58 & 0.0751 & 1.34 & 7.60 & 1.71 & 2.56 & 0.177 & 0.020 & 1.42 & 0.59 \\[0.5em] 
33000 & 8.30 & -13.96 & 0.1696 & 1.31 & 6.03 & 2.18 & 2.41 & 0.120 & 0.034 & 1.20 & 0.49 \\ 
33000 & 7.70 & -13.12 & 0.0736 & 1.30 & 7.69 & 1.51 & 2.56 & 0.182 & 0.016 & 1.39 & 0.58 \\ 
    \hline
\label{params2}
\end{tabular}
\end{minipage}
\end{table*}

Moreover, even in the formally unstable part of the convection zone, where
$F_{\rm kin}$ is small in magnitude compared to $F_{\rm conv}$ (around
$\log T\la 4.2$, i.e.\ where the largest velocities occur), non-local
effects remain important. The reason for this result is that in the
governing equation for the velocity
field in the non-local model, the third order moment directly related
to $F_{\rm kin}$, $\overline{q^2 w}=2 F_{\rm kin}/\rho$, appears under
a divergence 
(eq.~1 in \citealt{Kupka99}, eqs.~19a, 36a, and 51a in CD98). Thus,
the rapid variation of $\overline{q^2 w}$ can cause its divergence to
be large even if the third order moments themselves are small (e.g. as
compared to the convective flux). This divergence couples the velocity
field with the vertical structure of the model in an interesting
manner: fluid elements having differing degrees of partial ionisation
are transported to and from neighbouring layers. The ionisation
and recombination work and the ionisation energy contained in the gas
change the nature of the flow. They hence play an important role in the
non-local nature of convection driven by ionisation zones \citep[for
the related case of numerical simulations of solar granulation
see][]{Rast93}.

As a consequence, for models of both `early' white dwarfs and early A-stars
we find that their convection zones can be inefficient in the sense of
having small convective and kinetic energy fluxes while still having
quite large velocity fields, as seen in Fig.~\ref{models}, and that such
large velocities can have significant observational and theoretical
consequences. For A-stars, the studies of line profiles discussed in
\citet{Landstreet98} provide a strong observational indication for
large velocity fields with a large asymmetry between up- and downflows
in these objects.  Although such observational indicators are not yet
available for white dwarfs, the similarity of non-local models for
both A-stars and white dwarfs, which is supported independently by the
numerical simulations of \citet[ see also
Section~\ref{Sec_numsim}]{Freytag95}, suggests that white dwarf
envelopes can have significant convective velocities despite having a
nearly radiative temperature structure. Thus, the non-local effects,
far from being unimportant, manifest themselves in relatively large
convective velocities, which in turn can affect the formation and
shape of spectral lines in the atmosphere, the diffusion/gravitational
settling of elements in the envelope, as well as the way in which any
pulsations present in the star interact with the convection zone.

\subsection{DB models}  
\label{db}

For the DB's, we have calculated models with temperatures between
28,000~K and 35,000~K. The analogous results to the DA case are given
in Fig.~\ref{models_DB} and in Table~\ref{params2}.  Besides the
general trend that convection becomes more efficient as the
temperature decreases, we see that these results are actually quite
similar to what we found for the DA's. For instance, the maximum
convective velocities of both are in the range 5--6~km/s, their flux
overshooting is $\sim$~1.25~$H_P$, and their velocity overshooting is
$\sim$~2.5~$H_P$. One difference between the DA and DB models is the
remarkable increase in the photospheric velocity seen in
Fig.\ref{models_DB}b for the coolest model (28,000~K). This is caused
by additional driving provided by the partial ionisation of \hei. At
such low optical depths ($\tau \sim 2/3$) this ionisation zone cannot
alter the energy transport, but interestingly, at least within the
context of the present, non-local model, it is able to leave a very
clear feature in the photospheric surface velocity field. For higher
effective temperatures, this ionisation zone moves to even smaller
optical depths and its influence on convection vanishes completely.

The other principal differences between the DA and DB models turn out
to be in the depth of their convection zones and in their convective
efficiencies, as parametrised by $\alpha_{\rm eff}$: the DB's have
systematically lower values of $\alpha_{\rm eff}$, and for a given
convective flux the DB's have convection zones which are an order of
magnitude more massive than those of the DA's. As it turns out, these
two results are consistent with each other. Since the \heii\ 
convection zone is deeper than the \hi\ zone, the MLT eddies contain
denser material which has a higher heat capacity and is optically thick,
allowing them to transport the same fraction of the total flux with
a smaller value of $\alpha$, i.e., more efficiently.

\begin{figure}
\includegraphics[width={\columnwidth}]{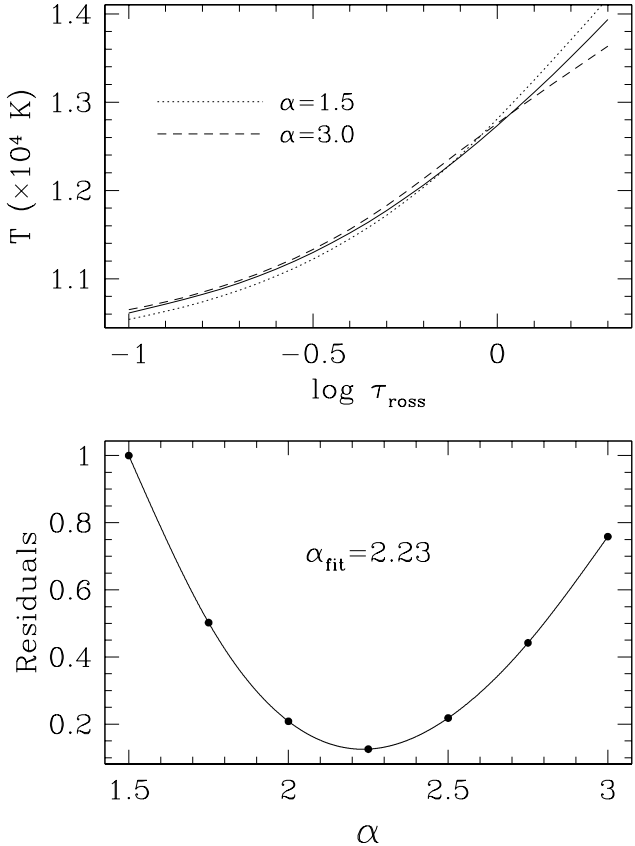}
\caption{Upper panel: The temperature structure of the photosphere of the
  \teff=12,200~K, $\log g$=8.0 model as a function of optical depth of
  the non-local model (solid curve) compared to that of local MLT
  models (ML1) having $\alpha=1.5$ (dotted curve) and $\alpha=3.0$
  (dashed curve). Lower panel: The residuals from a least squares fit
  as a function of $\alpha$ over this range. The best fit is achieved
  for $\alpha=$2.23; for ML2, we find $\alpha=$0.89.}
\label{alphafit}
\end{figure}

\section{Comparison with MLT}  
\label{mlt}

There are many different ways in which a comparison between the
Reynolds stress model and local MLT can be made. In
Section~\ref{results}, we chose to match the maximum of the convective
flux in the two models. Since the maxima occur at large optical
depths, we are in some sense matching the models in a `deep' region.
Alternatively, we could choose to do the comparison in the
photosphere, since this has relevance for model atmosphere fits. In
this section, we compare the temperature structure of the photospheres
of the local and non-local models; a comparison of the velocity fields
is not informative since the local models produce convection zones
which are always too narrow for a given convective flux. In addition,
the velocity fields used in model atmosphere fits are not usually
derived from an underlying convection model but are treated as a free
parameter (`microturbulence'), whereas the temperature profiles used
in the fits {\em are} computed using a model of the convection.

Since the temperature structure is most dependent on the convective
prescription for models with the largest photospheric convective
fluxes, we choose our lowest temperature models for the comparisons.
In Fig.~\ref{alphafit} we show the results of such a fitting procedure
for our \teff=12,200~K, $\log g$=8.0 DA model. In the upper panel we
plot the temperature as a function of optical depth. We see that the
non-local model (solid curve) is bracketed by the local MLT models
(ML1) having $\alpha=1.5$ (dotted curve) and $\alpha=3.0$ (dashed
curve).  In the lower panel, we plot the results of a least squares
fit of the MLT profiles to the non-local profile, as a function of
$\alpha$. The best fit is achieved for a value of $\alpha=2.23$. This
is in broad agreement with \citet{Koester94}, who find that
$\alpha=2.0$ yields model spectra which are in reasonable agreement
with the observations.  In addition, \citet{Ludwig94} found that the
temperature structure of their \teff=12,600~K hydrodynamic model could
be approximated with values of $\alpha$ between 1 and 2. For {\em our}
\teff=12,600~K model, we find it to be bracketed by $\alpha=1.0$ and
2.5 models, with a best fit value of $\alpha=1.86$, which is consistent
with their result.

Many of the current model atmosphere fits for white dwarfs are based
on a version of MLT in which radiative energy losses are somewhat
suppressed \citep{Bohm71}; this version, which is more efficient than
ML1 for a given value of $\alpha$, is often referred to as ML2
\citep[e.g.\ ][]{Tassoul90,Ludwig99}. For our \teff=12,200~K model, we
find that it is best fit with a value of $\alpha = 0.89$. For
our \teff=12,600~K model, we find that the best fit value of $\alpha$
is 0.73. This is in good agreement with \citet{Bergeron95}, who found
that their optical spectra of ZZ~Ceti's could be well fit by
ML2/$\alpha=1.0$ models, but that best fits to both the UV and optical
spectra were obtained with ML2/$\alpha=0.6$, which is the value they
adopted for their subsequent analysis.

For our DB models, we are unable to do such fits of the photospheric
convective efficiency since the models have so little convection in
the photosphere. Thus, the only results comparing our models with
local MLT models are the ones already given in Table~\ref{params2}.

\section{Comparison with numerical simulations}  \label{Sec_numsim}

\begin{table*}
\centering
\begin{minipage}{0.72\textwidth}
  \caption{Comparison of Reynolds stress model results to the numerical
    simulations of \citet{Freytag95}; all models have $\log g$=8.0.
    We have used $v_{\rm C}$ and $v_{\rm h}$ to denote the vertical
    and horizontal rms convective velocities, respectively.  Columns
    labelled {\sc rs} are the Reynolds stress results, and columns
    labelled {\sc f} are the results of Freytag's numerical
    simulations.}
  \begin{tabular}{ccccccccccccc}
    \hline
    \teff 
 & \multicolumn{2}{c}{$(F_{\rm C}/F_{\rm T})_{\rm max}$}
 & \multicolumn{2}{c}{$(F_{\rm C}/F_{\rm T})_{\tau=2/3}$}
 & \multicolumn{2}{c}{$(v_{\rm C})_{\rm max}$}
 & \multicolumn{2}{c}{$(v_{\rm C})_{\tau=2/3}$}
 & \multicolumn{2}{c}{$(v_{\rm h})_{\rm max}$} 
 & \multicolumn{2}{c}{$(v_{\rm h})_{\tau=2/3}$} \\
 (K)  & & & &
 & \multicolumn{2}{c}{(km~s$^{-1}$)} 
 & \multicolumn{2}{c}{(km~s$^{-1}$)} 
 & \multicolumn{2}{c}{(km~s$^{-1}$)} 
 & \multicolumn{2}{c}{(km~s$^{-1}$)}  \\
\hline & {\sc rs} & {\sc f} & {\sc rs} & {\sc f} & {\sc rs} & {\sc f} & 
{\sc rs} & {\sc f} & {\sc rs} & {\sc f}  & {\sc rs} & {\sc f} \\ 
    \hline  \\[-0.9em]
12200 & 0.440 & 0.385 & 0.119 & 0.036 & 6.43 & 3.95 & 6.22 & 1.95 & 6.95 & 6.38 & 5.42 & 6.26 \\
12600 & 0.126 & 0.182 & 0.042 & 0.043 & 5.33 & 3.31 & 4.16 & 2.22 & 4.31 & 6.01 & 3.53 & 5.80 \\
13000 & 0.064 & 0.063 & 0.036 & 0.020 & 5.40 & 2.63 & 4.22 & 1.86 & 4.34 & 4.14 & 3.53 & 3.95 \\
13400 & 0.035 & 0.008 & 0.019 & 0.006 & 5.50 & 1.39 & 4.41 & 1.14 & 4.42 & 2.23 & 3.66 & 1.72 \\
    \hline
\label{compare}
\end{tabular}
\end{minipage}
\end{table*}

In Table~\ref{compare}, we compare the results of our DA models to
those from the 2D simulations of \citet{Freytag95}. This comparison is
not completely fair due to the fact that different equations of state
have been used. Even so, we see that the agreement in terms of the
maximum convective fluxes is actually quite good, although the
photospheric fluxes do not agree as well.  In addition, both the
maximum and photospheric values of our vertical convective velocities
($v_C$) are systematically higher than his by a sizeable margin.  At
the present, we do not know if this signals an incompleteness in the
Reynolds stress model or an inadequacy in the 2D calculations, or
both.  For instance, \citet{Asplund00} found that while 2D simulations
in the Sun did reasonably well at reproducing the temperature
structure of the 3D simulations (and therefore the convective fluxes),
the 2D simulations systematically underestimated the magnitude of the
convective velocities by 10--20\%. In addition, the viscosity used for
the 2D the simulations may also lead to a reduction of the velocities.
Conversely, it is possible that some of the closures in our
implementation of the Reynolds stress model may be responsible for
this discrepancy \citep[e.g. see Fig.~5 of][]{Kupka03}. Finally,
perhaps surprisingly, we note that the discrepancy between our models
is smaller for the horizontal component of the convective velocity,
$v_{\rm h}$, than it is for the vertical component.

An interesting question concerns how the velocity field decays beneath
the convection zone. \citet{Freytag95} and \citet{Freytag96} claim
both numerical and theoretical evidence for an exponential decay with
depth of the turbulent velocity field; their position is given some
support by \citet{Ludwig03} for the case of overshooting into stellar
atmospheres, although he does not find this to be the
case for all the models examined. Our equations, on the other hand,
lead both analytically and numerically to an approximately linear
decay of the velocity field with depth. We claim that the correct
behaviour at the base of a convection zone is not yet known, since the
way in which one filters out the travelling waves from the truly
convective fluid motions has an important effect on the velocity field
inferred from the hydrodynamical simulations \citep{Ludwig03}. Waves,
on the other hand, are much less efficient in mixing a fluid than a
network of drafts and plumes associated with a `genuine' convective
velocity field. The resolution of this problem has an important
bearing on our understanding of diffusion in white dwarf envelopes,
since the depth to which convective fluid motions penetrate can
greatly affect the diffusion of chemical elements.

Finally, we mention a technical point concerning the Reynolds stress
model, which we examine in detail in an appendix. The formalism of
\citet{Canuto98}, which we employ, implicitly assumes for the purpose
of statistical averaging that the equation of state is that of an
ideal gas. This means that thermodynamic quantities are held constant
during averaging, i.e., $\overline{c_P\,w \theta}$ is replaced by $c_P
\, \overline{w \theta}$.\footnote{We note that in our subsequent
  implementation of the Reynolds stress equations that we \emph{do}
  use the actual values of $c_P$ and other thermodynamic quantities
  computed using the OPAL equation of state.}  This is not strictly
valid since convection zones usually coincide with regions which are
partially ionised, so that $c_P$ (and other thermodynamic quantities)
are not constants but functions of temperature and density. In
appendix~\ref{ideal}, we examine the effect of this approximation and
show that it leads to errors in the convective flux no larger than
10--20\% for the calculations we have done. Since this uncertainty is
less than that introduced by other aspects of the modelling, such as
different prescriptions for the third order moments \citep[see Fig.~5
of][ for the case of A-stars]{Kupka03}, we are justified in neglecting
this effect in the present analysis.  In addition, we show that in the
future the correction terms arising from relaxing this assumption can
be naturally included within the Reynolds stress formalism.

\section{Comparison with pulsation data} 

For pulsations in the DA and DB stars, we are in the fortunate regime
in which the convective turnover time is quite short ($\sim$1~s)
compared to the periods of the observed modes ($\sim$100~s). This
means that the convection zone responds to the pulsations in a
quasi-static manner, so the only input needed for the pulsation
calculations is a sequence of static convection zones computed for
different effective temperatures. Thus, true time-dependent solutions
for the convective region are not needed, which simplifies the
modelling considerably.

It has recently been shown
\citep{Brickhill91a,Brickhill91b,Goldreich99a} that the convection
zone plays a vital role in the driving of pulsations in the DAV stars,
and this is likely to be true of the DBV stars as well. Thus, the
observed blue edge and red edge of the instability strip \citep[see
][]{Bergeron95,Bergeron03} should contain information concerning the
dependence of the convective efficiency on \teff\ and $\log g$.

In addition, the observed nonlinearities in the lightcurves of the
DAV's are also believed to be due to the interaction of the convection
zone with the pulsations \citep{Wu01,Ising01}, so it should again be
possible to use the observations to constrain different models of
convection; this work is presently under way (Montgomery, in
preparation). Thus, the pulsating white dwarfs (both DAV's and DBV's)
offer a great deal of promise for learning about the physics of
convection.

\section{Conclusions}  

Using a fully non-local model of convection together with a realistic
equation of state and opacities, we have calculated envelope models
for stellar parameters appropriate for DA and DB white dwarfs.  We
find good agreement between our models and those obtained through
fitting white dwarf spectra, as well as good agreement with the
results of hydrodynamic simulations.

First, the maximum convective fluxes in our DA models compare
reasonably well with those found in 2D hydrodynamical simulations
\citep{Freytag95}. This result appears more impressive when taking
into account that a similar agreement was found for the case of
A-stars \citep{Kupka02} while MLT type models such as ML1 require a change
of the scale length parameter $\alpha$ by a factor of 4 (see already
\citealt{Freytag95}). On the other hand, the convective velocities (both
vertical and horizontal) differ from each other by a fairly
significant amount. At present we do not know whether this signals an
incompleteness in the Reynolds stress model or a limitation of the
2D simulations, or both. However, while our rms velocities and those
of Freytag (1995) are found to differ by up to a factor of two in the
convection zone, we emphasise the completely different results
obtained when using a local convection model such as ML1 or ML2. Among
other shortcomings, the latter underestimate the size of the
convectively mixed region by a factor of at least 10 in terms of mass
relative to our new results and even more when compared to those of
\citet{Freytag95}. Thus, while the predicted rms velocities at least
qualitatively agree when comparing the Reynolds stress formalism
applied in this paper with 2D numerical simulations, results from
local convection models such as MLT are fundamentally different from
the two. 

Second, given the widespread use of MLT in stellar and atmospheric
modelling, we have compared aspects of our models to MLT models.  We
find for our DA models that their photospheric temperature structure
is best approximated by local MLT models (ML2) having $\alpha$ between
$\sim\,$0.7 and $\sim\,$0.9. This is in agreement with the results of
\citet{Bergeron95}, who found that their model spectra matched both
the optical and UV data best for ML2/$\alpha$=0.6; for fitting the
optical data alone, ML2/$\alpha$=1.0 also provided a good fit to their
observations. Going beyond such fits, our non-local model offers the
promise of a unified model for self-consistently computing the
velocity field (`microturbulence') which is also required for such
model atmosphere fits, something which local models fail to do.

Since the onset of convection goes hand in hand with the onset of
pulsation in both the DA's and DB's, we have the opportunity to use
asteroseismology to sample their interior structure. First of all, the
dependence of the blue edge of the instability strip on \teff\ and
$\log g$ very likely depends on how the convective efficiency varies
with these parameters, so the observed blue edge can be used to
constrain theories of convection. In addition, recent calculations
have shown that the convection zones in these stars may be the main
source of the observed non-linearities in their lightcurves
\citep{Brickhill92a,Wu01,Ising01}, so it may be possible to use the
pulsations themselves to probe the structure of the convection zone.
In particular, from observations of a given pulsating star it may be
possible to infer the depth of the convection zone as a function of
the instantaneous effective temperature (Montgomery, in preparation).
This would provide important data for the current convection models.

From an astrophysical standpoint, white dwarf stars are crucial for
our understanding of the initial-final mass relation for stars as well
as providing an independent method for determining stellar and
Galactic ages. In addition, through asteroseismology, they can serve
as test beds for nuclear reaction rates \citep{Metcalfe03}, chemical
diffusion \citep{Montgomery01}, crystallisation
\citep{Winget97,Montgomery99a}, and neutrino emission
\citep{Obrien98}. In order to make maximum use of these inferences,
however, model atmospheres need to be applied to the observations of
individual white dwarfs in order to derive the stellar parameters
($M_{\star}$, \teff); depending upon the prescription which is adopted
for convection, these parameters can vary significantly \citep[e.g.
][]{Bergeron95}.  Thus, understanding convection in the context of
these stars is vital.

Except for a possible application to RR Lyr stars and Cepheids, the
results reported here and in \citet{Kupka02} represent the limits of
what we can treat using our present semi-implicit solver. We are
currently in the process of developing a fully-implicit solver, which
will allow us to treat thicker convection zones having larger thermal
timescales, and, hence, lower effective temperatures. In addition to
allowing us to compute cooler models with deeper convection zones for
the DA and DB white dwarfs, and the A and F-stars, we would eventually
like to study the convection zone of the Sun. This would be an
excellent test of the model since a wealth of high-quality data
already exists for the solar convection zone. On the same basis,
calculations of mixing and overshooting in stellar convective cores
may be an even more important application. While the quality of data
for probing the effects of convective core overshooting cannot match
some of the solar observational results, there are no extra
complications as introduced by the large temperature fluctuations and
non-grey radiative transfer near the solar surface, which are
unaccounted for by the present model but are expected to alter basic
quantities such as the derived depth of the solar convection zone.
The current accuracy of evolution models of intermediate and high mass
stars is still most severely limited by the coarse treatment of
convective heat transfer and mixing. At least part of the solution to
this problem might be achieved by using a model similar to the one
applied in this paper.

\section*{Acknowledgments}

We thank Dr.\ B.~Freytag for providing us with results from his
simulations and for stimulating discussions, and we also thank Prof.
V. Canuto for his insightful comments on our work. This research was
supported by the UK Particle Physics and Astronomy Research Council
through grants PPA/G/O/2002/00498 and PPA/G/O/1998/00576.

\appendix

\section{Estimate of errors arising from the ideal gas assumption}

\label{ideal}

In this appendix, we give an assessment of the uncertainties
introduced by the ideal gas assumption in the Reynolds stress model
formalism of \citet{Canuto98} we mentioned in Section~\ref{Sec_numsim}.
Essentially, this has meant that the specific heats, $c_P$ and $c_V$,
were treated as constants when computing ensemble averages. To
partially correct for this, we have used a realistic equation of
state to compute the values of $c_P$ and $c_V$ as functions of
$\rho$ and $T$ where they occur in the relevant equations. However,
we are necessarily missing any cross terms (moments) that may have
arisen when they were averaged against other fluctuating quantities
to derive the moment equations themselves.

To provide an estimate of the errors which this may introduce, we
consider the convective flux, which is actually the flux of enthalpy,
$h$. From \citet{Canuto97}, equation (19b), we have
\[
F^C_j = \overline{\rho h u_j},
\]
which, in the absence of a mean flow, we approximate as
\begin{equation}
F^C_j = \overline{\rho} \, \overline{h' u_j'},
\label{fc}
\end{equation}
in the Boussinesq sense. Here, $u_j$ is the $j$-th component of the
turbulent velocity, and the overbar denotes an ensemble average. In
the present context, we do not consider density-weighted averages,
which are more convenient when dealing with a fully compressible flow.
In what follows, all variables have been separated into an average and
a fluctuating component with respect to ensemble averages.  For
example, if $X$ represents a given fluctuating quantity (e.g.
temperature, velocity, etc.), we write
\begin{equation}
  X = \overline{X} + X',
\end{equation}
where
\begin{equation}
  \overline{X'} = 0.
\end{equation}

\begin{figure*}
\includegraphics[width=1.00\columnwidth]{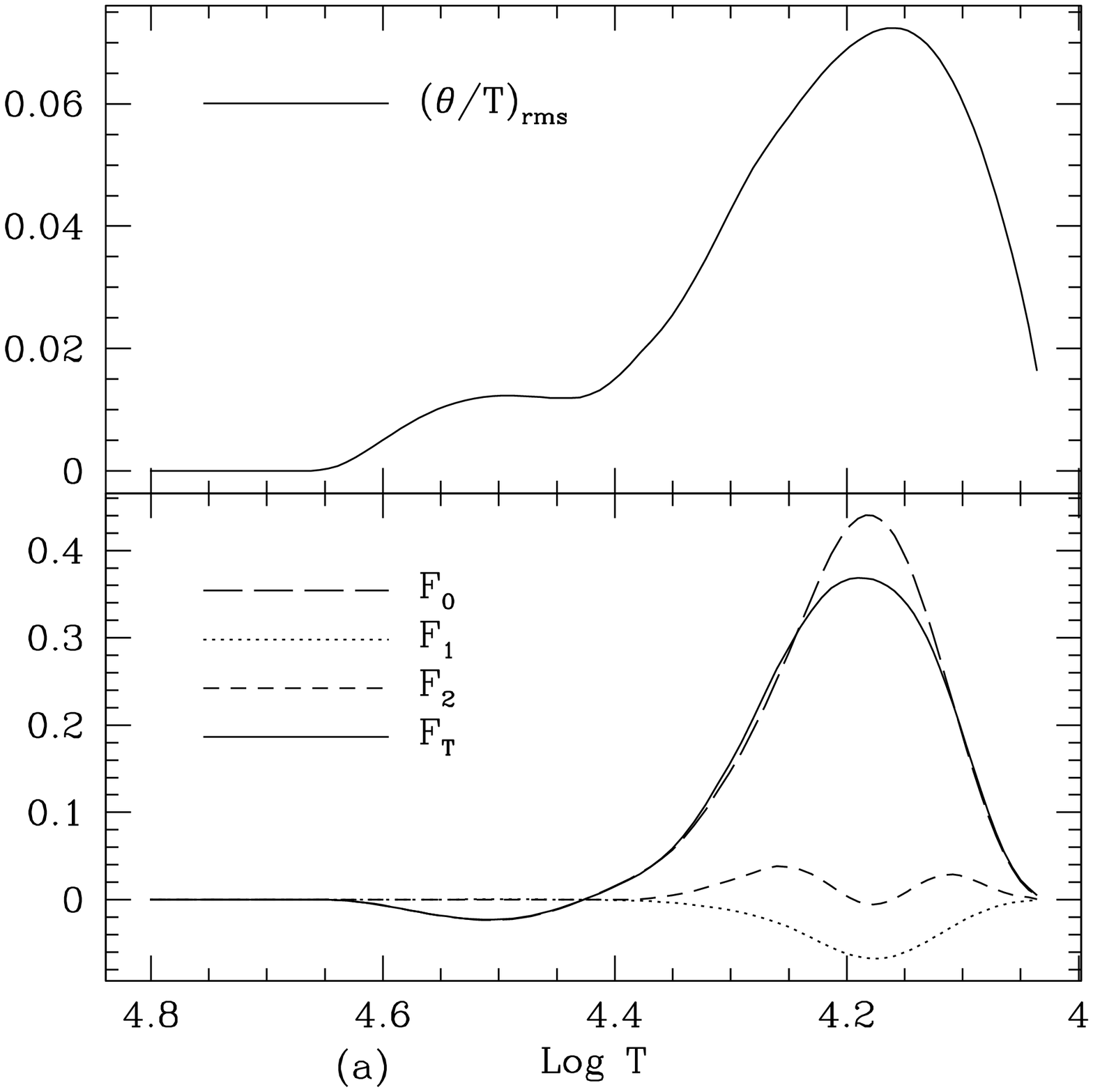}
\hfill
\includegraphics[width=1.00\columnwidth]{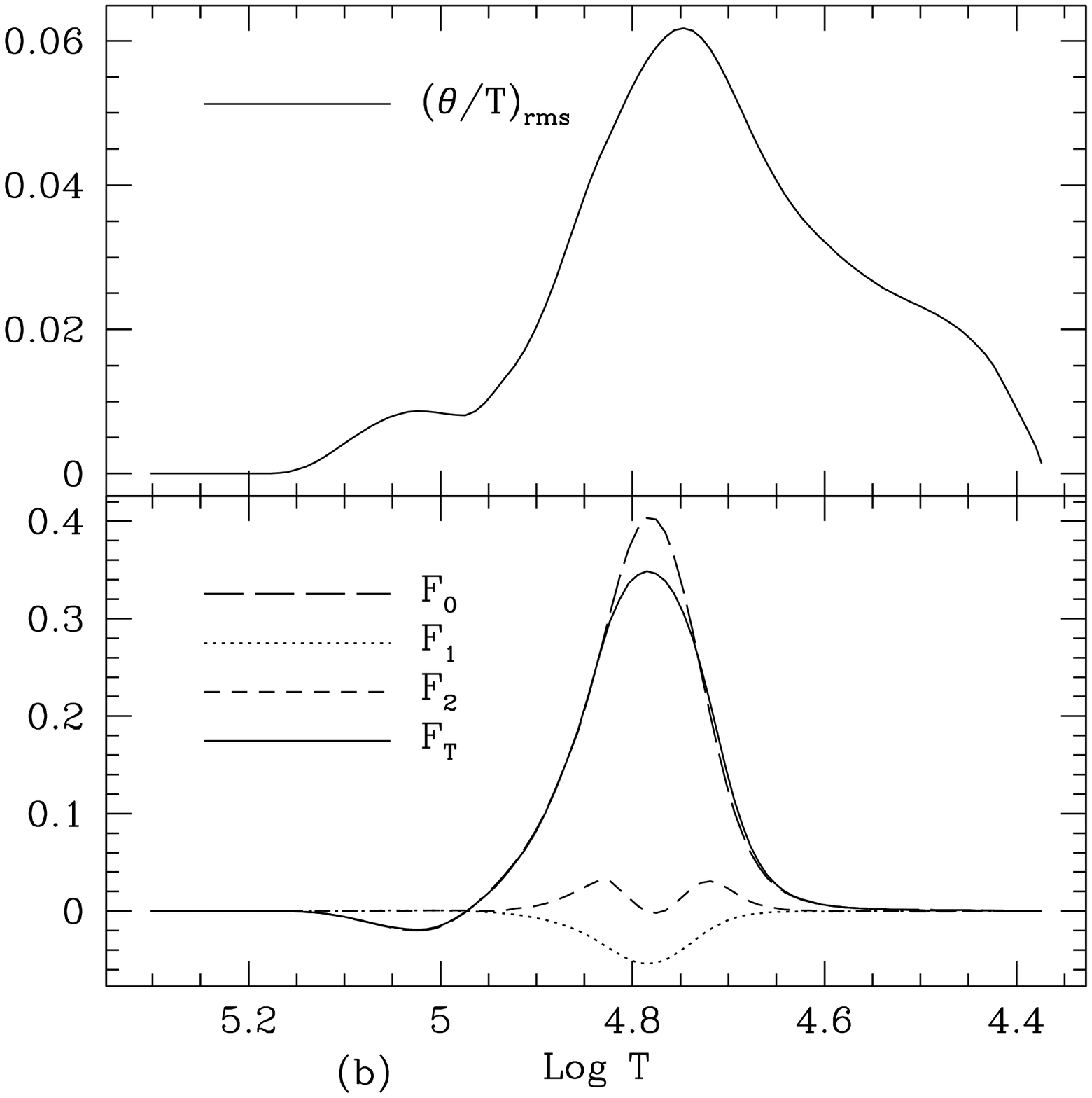}
\caption{The rms fractional temperature fluctuation (upper panel)
  and the convective fluxes (lower panel) for a \teff=12200~K DA model
  (a) and a \teff=28,000~K DB model (b).  The fluxes shown in the
  lower panel are those given in equation~(\protect\ref{fterms}), with
  the sum of these, $F_{\rm T}$, indicated by the solid curve.
\label{flux_corr}
}
\end{figure*}

In the work of Canuto \citep{Canuto97,Canuto98} the assumption has
been made that the gas is ideal, specifically that $h=c_P T$, with
$c_P$ being constant.  Here, we consider $h$ to be a function of $T$
and $P$, i.e., $h= h(T,P)$. If we assume that $T' \ll \overline{T}$,
$P' \ll \overline{P}$, we have
\begin{eqnarray}
h & = & h(\overline{T}+T',\overline{P}+P') \\
   & = & h(\overline{T},\overline{P}) + h_T\, T' + h_P\, P' + 
   \frac{1}{2} h_{TT} \, T'^2 + O(T'P',P'^2),
   \label{hexp}
\end{eqnarray}
where
\[
\begin{array}{ccccl}
h_T & \equiv & \left(\frac{\partial h}{\partial T}\right)_P
    & = & c_p \\
h_P & \equiv & \left(\frac{\partial h}{\partial P}\right)_T
    & = & (1-v_T)/\rho \\
h_{TT} & \equiv & \left(\frac{\partial^2 h}{\partial T^2}\right)_P
    & = & \left(\frac{\partial c_p}{\partial T}\right)_{\rho} - 
    \frac{\rho v_T}{T} \left(\frac{\partial c_p}{\partial \rho}\right)_T, \\
\end{array}
\]
and
\[
v_T \equiv \frac{\chi_T}{\chi_{\rho}} 
 = - \left(\frac{\partial \ln \rho}{\partial \ln T}\right)_P,
\]
and each of the quantities $h_T, h_P, h_{TT}, c_p, \nu_T$ is evaluated
at $\overline{T}, \overline{P}, \overline{\rho}$ both here and in what
follows. Since in the ideal gas case the first order temperature term
is already present, and since we wish to go one order beyond this in
the temperature and pressure fluctuations, we take the expansion to
second order in temperature and first order in pressure.  As we will
see later, the first-order pressure corrections and the second-order
temperature corrections are indeed of similar magnitude.

Taking the ensemble average of equation~(\ref{hexp}), we find that
\[
\overline{h} = h(\overline{T},\overline{P}) + \frac{1}{2} h_{TT}
  \overline{T'^2}.
\]
Using $h'=h-\overline{h}$, we can substitute the above result into
equation~(\ref{fc}) for the convective flux. Writing $w$ and $\theta$
for the turbulent velocity and temperature perturbations, (i.e.,
$\theta \equiv T'$, $w \equiv u_j$), we find that
\[
F_C  =  \overline{\rho} \,h_T\, \overline{w \theta} + 
        \overline{\rho} \,h_P \,\overline{w P'} + 
\frac{1}{2} \overline{\rho} \,h_{TT} \,\overline{w (\theta^2 -\overline{\theta^2})} \,+
        \,\dots 
\]
Since $h_T = c_P$, the first term is the zeroth order term which we
have been calling the convective flux, and the other terms are the
corrections to it. Let us denote these terms by
\begin{eqnarray}
F_{0} & = & \overline{\rho} \,c_p \,\overline{w \theta} \nonumber \\
F_{1} & = & \overline{\rho} \,h_P \,\overline{w P'} \nonumber \\
F_{2} & = & 
  \frac{1}{2} \overline{\rho} \,h_{TT} \,\overline{w (\theta^2-\overline{\theta^2})} \nonumber \\
      & = & \frac{1}{2} \overline{\rho} \,h_{TT} \,\overline{w \theta^2}.
\label{fterms}
\end{eqnarray}
The quantity $\overline{w P'}$ is not directly calculated in the
present version of the code, although it may be approximated by
closures such as those found in \citet{Canuto92,Canuto97}. For our
present purposes, we use the relation
\[
  P' = \left(\frac{\partial P}{\partial T}\right)_{\rho} T'
      = \left(\frac{\partial P}{\partial T}\right)_{\rho} \theta,
\]
to recast $\overline{w P'}$ in terms of $\overline{w \theta}$, which
would be exact in the incompressible limit.

In Fig.~\ref{flux_corr}, we examine the relative sizes of these terms
for two different white dwarf models: a 12,200~K DA model (a) and a
28,000~K DB model (b).  In the upper panels we show the rms fractional
temperature fluctuations as taken from our converged non-local models;
we note that these fluctuations are everywhere less than 8\%,
justifying our treatment of them as small perturbations. In the lower
panel, we plot the leading order flux, $F_0$ (long-dashed curve), the
first order pressure correction to it, $F_1$ (dotted curve), the
second order temperature correction, $F_2$ (short dashed curve), and
the sum of all of these, $F_{\rm T}$ (solid curve).  We see that while
these correction terms are non-negligible, they would lead to
fractional corrections to $F_0$ of only about 15\%. In addition, these
corrections appear to be large only in the central part of the
convection zone; outside of this region, the convective (enthalpy)
flux is essentially given by $F_0$.

\begin{figure}
\includegraphics[width=\columnwidth]{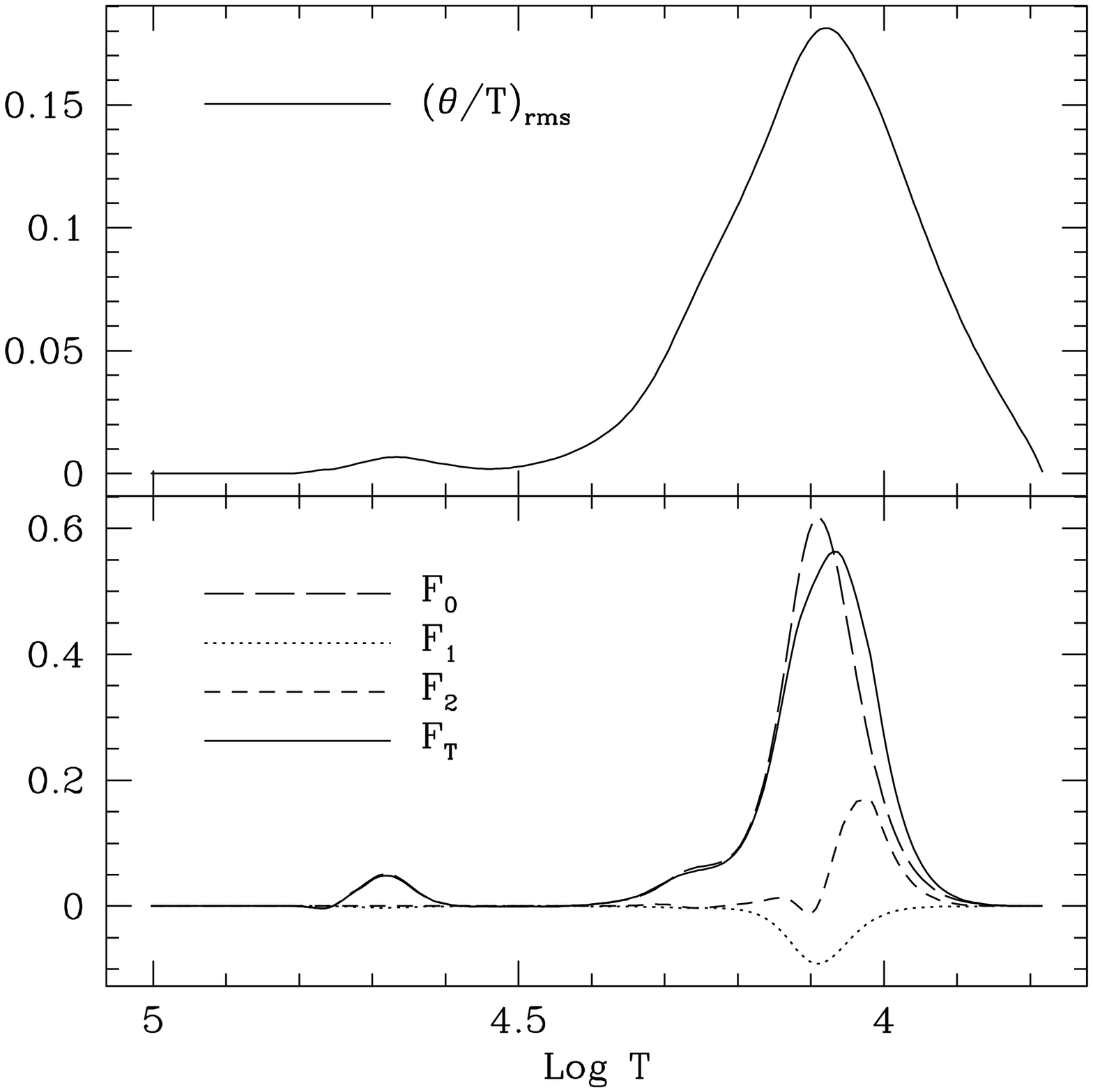}
\caption{Same as Fig.~\protect\ref{flux_corr} but for an A-star model having
  \teff=7200~K and $\log g = 4.4$.}
\label{flux_corr2}
\label{lastpage}
\end{figure}

Finally, in Fig.~\ref{flux_corr2}, we make a similar plot for an
A-star model (\teff=7200~K, $\log g=4.4$) from our previous paper
\citep{Kupka02}. In contrast to the white dwarf case, we see that
there are two different convectively unstable regions, corresponding
to \heii\ ionisation ($\log T \sim 4.7$) and \hi\ ionisation ($\log T
\sim 4.1$). In the deeper \heii\ zone, we see that these correction
terms are negligible. In the \hi\ zone, these terms are larger, driven
by the fairly large temperature fluctuations of up to $\sim 18$\%.  We
see that the main effect is to shift the maximum of the convective
flux slightly outward and to decrease its magnitude; even so, this
decrease in the maximum convective flux is only about 10\%.  As in the
white dwarf case, we see that these corrections are only significant
within the formal convection zone and not in the overshooting regions.

We note that for both Figs.~\ref{flux_corr} and \ref{flux_corr2} the
models have not been reiterated, i.e., the converged model does not
include the corrections to $F_0$. However, as the difference between
$F_0$ and $F_{\rm T}$ is no more than 15\% and as there is a negative
feedback, the figures are sufficient to demonstrate the size of the
effect. Self-consistent models for the cases shown in Fig.~\ref{flux_corr}
which are based on the full equation~(\ref{fterms}) for $F_{\rm T}$
are expected to have a larger convective driving and thus a maximum
in $F_{\rm T}$ larger than the one plotted, though still smaller than
$F_0$. The case shown in Fig.~\ref{flux_corr2} is not quite so simple,
but a self-consistent solution can be expected to yield a flux which
lies between $F_T$ and $F_0$, regardless of which one is larger.

The additional correction terms which we have derived above (i.e.,
those in equation~\ref{fterms}) pose no problem for the Reynolds
stress approach since these new terms involve moments which are
already calculated in the current formalism ($\overline{w \theta},
\overline{w \theta^2}$ in the current paper, and eventually
$\overline{w P'}$). Thus, even if it turns out that these terms are
more important for thicker/cooler convection zones, it should not
prevent us from extending the Reynolds stress approach further into
this regime.

Finally, we note that the magnitude of the uncertainties in
equation~(\ref{hexp}) is of order $\sim\,10\,$\%, which is actually
less than that introduced by other aspects of the modelling, such as
different prescriptions for the third order moments \citep[see Fig.~5
of][ for the case of A-stars]{Kupka03}. Thus, in the present analysis
we are justified in neglecting this effect.

\end{document}